\documentclass[prl,twocolumn,superscriptaddress,preprintnumbers,amsmath,amssymb]{revtex4}
\usepackage{graphicx}
\usepackage{amsmath}
\usepackage{latexsym}
\usepackage{dcolumn}
\usepackage{bm}
\usepackage{float}
\usepackage{graphicx,here}
\usepackage{color}

\emergencystretch=\maxdimen
\begin{document}

\title{\bf Observation of one-dimensional Dirac fermions in silicon nanoribbons}

\affiliation{Institute of Physics, Chinese Academy of Sciences, Beijing 100190, China}
\affiliation{Beijing Academy of Quantum Information Sciences, Beijing 100193, China}
\affiliation{Key Lab of Advanced Optoelectronic Quantum Architecture and Measurement (MOE), Beijing Key Lab of Nanophotonics Ultrafine Optoelectronic Systems, and School of Physics, Beijing Institute of Technology, Beijing 100081, China}
\affiliation{Hiroshima Synchrotron Radiation Center, Hiroshima University, 2-313 Kagamiyama, Higashi-Hiroshima 739-0046, Japan}
\affiliation{School of Physical Sciences, University of Chinese Academy of Sciences, Beijing 100049, China}
\affiliation{Songshan Lake Materials Laboratory, Dongguan, Guangdong 523808, China}

\author{Shaosheng Yue}
\affiliation{Institute of Physics, Chinese Academy of Sciences, Beijing 100190, China}
\affiliation{School of Physical Sciences, University of Chinese Academy of Sciences, Beijing 100049, China}
\author{Hui Zhou}
\affiliation{Institute of Physics, Chinese Academy of Sciences, Beijing 100190, China}
\affiliation{School of Physical Sciences, University of Chinese Academy of Sciences, Beijing 100049, China}
\author{Ya Feng}
\affiliation{Beijing Academy of Quantum Information Sciences, Beijing 100193, China}
\author{Yue Wang}
\affiliation{Key Lab of Advanced Optoelectronic Quantum Architecture and Measurement (MOE), Beijing Key Lab of Nanophotonics Ultrafine Optoelectronic Systems, and School of Physics, Beijing Institute of Technology, Beijing 100081, China}
\author{Zhenyu Sun}
\affiliation{Institute of Physics, Chinese Academy of Sciences, Beijing 100190, China}
\affiliation{School of Physical Sciences, University of Chinese Academy of Sciences, Beijing 100049, China}
\author{Daiyu Geng}
\affiliation{Institute of Physics, Chinese Academy of Sciences, Beijing 100190, China}
\affiliation{School of Physical Sciences, University of Chinese Academy of Sciences, Beijing 100049, China}
\author{Masashi Arita}
\affiliation{Hiroshima Synchrotron Radiation Center, Hiroshima University, 2-313 Kagamiyama, Higashi-Hiroshima 739-0046, Japan}
\author{Shiv Kumar}
\affiliation{Hiroshima Synchrotron Radiation Center, Hiroshima University, 2-313 Kagamiyama, Higashi-Hiroshima 739-0046, Japan}
\author{Kenya Shimada}
\affiliation{Hiroshima Synchrotron Radiation Center, Hiroshima University, 2-313 Kagamiyama, Higashi-Hiroshima 739-0046, Japan}
\author{Peng Cheng}
\affiliation{Institute of Physics, Chinese Academy of Sciences, Beijing 100190, China}
\affiliation{School of Physical Sciences, University of Chinese Academy of Sciences, Beijing 100049, China}
\author{Lan Chen}
\affiliation{Institute of Physics, Chinese Academy of Sciences, Beijing 100190, China}
\affiliation{School of Physical Sciences, University of Chinese Academy of Sciences, Beijing 100049, China}
\affiliation{Songshan Lake Materials Laboratory, Dongguan, Guangdong 523808, China}
\author{Yugui Yao}
\affiliation{Key Lab of Advanced Optoelectronic Quantum Architecture and Measurement (MOE), Beijing Key Lab of Nanophotonics Ultrafine Optoelectronic Systems, and School of Physics, Beijing Institute of Technology, Beijing 100081, China}
\author{Sheng Meng\footnote[1]{smeng@iphy.ac.cn}}
\affiliation{Institute of Physics, Chinese Academy of Sciences, Beijing 100190, China}
\affiliation{School of Physical Sciences, University of Chinese Academy of Sciences, Beijing 100049, China}
\author{Kehui Wu\footnote[2]{khwu@iphy.ac.cn}}
\affiliation{Institute of Physics, Chinese Academy of Sciences, Beijing 100190, China}
\affiliation{School of Physical Sciences, University of Chinese Academy of Sciences, Beijing 100049, China}
\affiliation{Songshan Lake Materials Laboratory, Dongguan, Guangdong 523808, China}
\author{Baojie Feng\footnote[3]{bjfeng@iphy.ac.cn}}
\affiliation{Institute of Physics, Chinese Academy of Sciences, Beijing 100190, China}
\affiliation{School of Physical Sciences, University of Chinese Academy of Sciences, Beijing 100049, China}

\date{\today}

\clearpage

\begin{abstract}
\section{Abstract}
Dirac materials, which feature Dirac cones in reciprocal space, have been one of the hottest topics in condensed matter physics in the last decade. To date, two- and three-dimensional Dirac fermions have been extensively studied while their one-dimensional counterparts are rare. Recently, Si nanoribbons (SiNRs), which are composed of alternating pentagonal Si rings, have attracted intensive attention. However, the electronic structure and topological properties of SiNRs are still elusive. Here, by angle-resolved photoemission spectroscopy and scanning tunneling microscopy/spectroscopy measurements, first-principles calculations, and tight-binding model analysis, we demonstrate the existence of 1D Dirac fermions in SiNRs. Our theoretical analysis shows that the Dirac cones derive from the armchair-like Si chain in the center of the nanoribbon and can be described by the Su-Schrieffer-Heeger model. These results establish SiNRs as a platform for studying the novel physical properties in 1D Dirac materials.
\\ \hspace*{\fill} \\
{\bf Keywords:} 1D Dirac fermions, Si nanoribbon, ARPES, SSH model, DFT calculations
\end{abstract}
\maketitle

\section{I. Introduction}

Dirac materials represent a special class of quantum matter hosting linearly dispersing Dirac cones \cite{HasanMZ2010,Wehling2014,WangJ2015,VafekO2014}. Because of the existence of Dirac cones, the low-energy excitation of Dirac materials behaves as massless Dirac fermions, which is the origin of various exotic properties, such as suppression of backscattering, high carrier mobility, and the quantum spin Hall effect. The first material confirmed to host Dirac fermions was graphene \cite{GeimAK2007,CastroNeto2009}. Later, several two-dimensional (2D) graphene-like materials, such as silicene and borophene, were predicted or confirmed to host Dirac fermions \cite{VogtP2012,FengB2017,FengB2019}. In addition to 2D materials, Dirac fermions have also been observed in three-dimensional (3D) materials, including Dirac semimetals \cite{ArmitageNP2018} and strong topological insulators \cite{HasanMZ2010}. Recently, the trend towards miniaturization of quantum devices has aroused much research interest in one-dimensional (1D) materials ({\it e.g.}, nanotubes, nanoribbons, and nanowires). However, to date, it is still challenging to realize Dirac fermions in 1D materials, although 1D Dirac-like bands have been observed in several 2D and 3D systems \cite{Gibson2014,KaneCL2005,MattesL2014,YangTY2020,LuJL2017,FengB2017nc,WellsJW2009,LeuenbergerD2013,BianchiM2015}, including the quantum spin Hall states at the boundaries of 2D topological insulators \cite{KaneCL2005,MattesL2014}, the confined metallic surface in van der Waals materials \cite{YangTY2020}, the Dirac nodal lines in 2D materials \cite{LuJL2017,FengB2017nc}, and the metallic states on the vicinal surface of bismuth \cite{WellsJW2009,LeuenbergerD2013,BianchiM2015}. 

As one of the simplest model to describe 1D topological materials, the Su-Schrieffer-Heeger (SSH) model \cite{SuW1979} has attracted great attention because of the rich physics in such a simple structure. To date, the SSH Hamiltonian have been realized in various systems, including cold atoms \cite{GaneshanS2013,MeierEJ2016}, wave guide arrays \cite{Longhi2013,KeS2019}, Peierls-dimerized atomic chains \cite{CheonS2015,HudaMN2020}, and graphene nanoribbons \cite{CaoT2017,GroningO2018,RizzoDJ2018,SunQ2020}. Angle-resolved photoemission spectroscopy (ARPES) is a powerful technique to study the topological electronic structures in condensed matter systems. However, ARPES studies of the SSH model are quite challenging because of the difficulty in the synthesis of large-scale, well-aligned 1D materials.

In this work, we show that Si nanoribbons (SiNRs) are ideal 1D materials hosting Dirac fermions and the topological properties can be described by the SSH model. As a 1D polymorph of Si, SiNRs have attracted intensive research interest in the last decades \cite{LeandriC2005,SahafH2007,ValbuenaMA2007,Padova2008,Padova2010}. Experimental synthesis of SiNRs on Ag(110) is easy and was realized more than a decade ago \cite{LeandriC2005}. However, the structure of SiNRs has been debated for over a decade, and various structure models have been proposed, such as silicene nanoribbons with or without edge reconstructions \cite{Padova2010,KaraA2010,LianC2012,TchalalaMR2014,FengB2016}. Recently, the structure of SiNRs has been confirmed as alternating pentagons by various methods, including first-principles calculations \cite{CerdaJI2016}, grazing incidence X-ray diffraction \cite{PrevotG2016}, X-ray photoelectron diffraction \cite{EspeterP2017}, atomic force microscopy \cite{ShengS2018}, tip-enhanced Raman spectroscopy \cite{ShengS2018}, and optical spectroscopy \cite{HogenC2018}. Regarding the electronic structures of SiNRs, early angle-resolved photoemission spectroscopy (ARPES) studies claimed the existence of a gapped Dirac cone \cite{Padova2010}. However, those experiments were performed with only 78-eV photons \cite{Padova2010}, which makes it difficult to disentangle the contribution from SiNRs and the Ag(111) substrate. Later, the Dirac-like bands in ARPES spectra were reinterpreted as a band folding effect induced by the periodicity of the nanoribbons \cite{GoriP2013}. To date, it is generally believed that SiNRs are metallic in nature \cite{FengB2016,RonciF2010,HiraokaR2017}, while their electronic band structures and topological properties are still elusive.

Here, we study the electronic structures and topological properties of SiNRs/Ag(110) by ARPES measurements and theoretical calculations. We observe a 1D Dirac cone with a Fermi velocity as high as 1.3$\times$10$^6$ m/s, which is slightly higher than that of graphene ($\sim$1.0$\times$10$^6$ m/s). The existence of the 1D Dirac cone is supported by our first-principles calculations. Using a simple tight-binding model, we reveal that the Dirac bands originate from an armchair-like Si chain in the center of the nanoribbon. These results might stimulate further research interest in studying the exotic properties of 1D Dirac materials.

\section{II. Experimental Results}

\begin{figure}[tbh]
\centering
\includegraphics[width=8 cm]{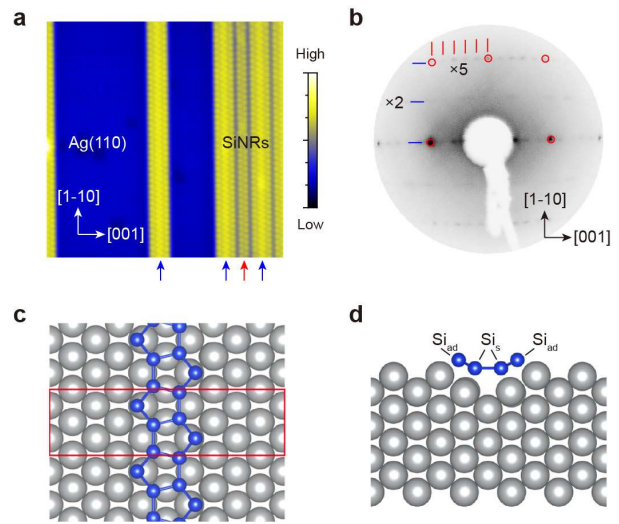}
\caption{{\bf STM and LEED characterization.}(a) STM topographic image of SiNRs on Ag(110). The substrate temperature during growth was approximately 450 K. Red and blue arrows indicate single- and double-strand SiNRs. (b) LEED pattern of SiNRs on Ag(110), showing a 5$\times$2 superstructure with respect to the 1$\times$1 lattice of Ag(110). Red circles indicate the spots of the Ag(110) substrate. (c) and (d) Top and side views of the relaxed structure model of SiNRs on Ag(110), respectively. The blue and gray balls represent Si and Ag atoms, respectively. The red rectangle indicates a unit cell of SiNRs/Ag(110). The two different types of Si atoms are marked as Si$_s$ and Si$_{ad}$.}
\end{figure}

Depending on the growth temperature, Si can form two types of nanoribbons on Ag(110). At room temperature, only single-strand SiNRs ($\sim$0.8 nm wide) exist on the Ag(110) surface. When the growth temperature is increased to $\sim$450 K, the surface is dominated by double-strand SiNRs ($\sim$1.6 nm wide), as shown in Fig. 1(a). Compared to single-strand SiNRs, double-strand SiNRs are perfectly aligned, and their lengths are limited only by the step edges and defects on the Ag(110) surface. Therefore, we focus on double-strand SiNRs because they are more suitable for ARPES measurements. A low-energy electron diffraction (LEED) pattern of the double-strand SiNRs is shown in Fig. 1(b). We observe a 5$\times$2 superstructure with respect to the 1$\times$1 lattice of Ag(110), in agreement with previous results \cite{SahafH2007}. Notably, the LEED patterns are elongated along the [001] direction of Ag(110) (perpendicular to the SiNRs). This occurs because the coverage of Si is less than one monolayer and there is no perfect periodicity in the [001] direction for the SiNRs, as evidenced by the STM image in Fig. 1(a). According to recent theoretical and experimental works, SiNRs consist of alternating Si pentagons \cite{CerdaJI2016,PrevotG2016,EspeterP2017,ShengS2018}, as shown in Figs. 1(c) and 1(d). The unique pentagonal nature may endow SiNRs with exotic physical properties.

\begin{figure*}[htb]
\centering
\includegraphics[width=15 cm]{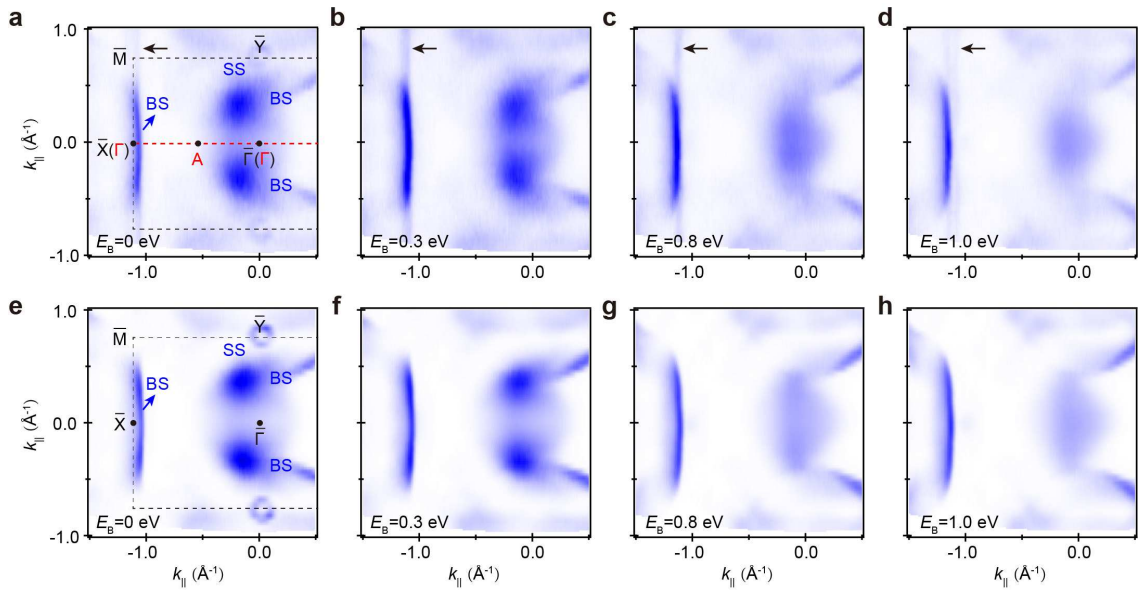}
\caption{{\bf ARPES measurements of the constant energy contours.} (a)-(d) ARPES intensity plots of SiNRs on Ag(110) at different binding energies: 0, 0.3, 0.8, and 1.0 eV. The incident photon energy was 35 eV. The substrate temperature during growth was approximately 450 K. (e)-(h) ARPES intensity plots of the pristine Ag(110) surface. The black and red dashed lines in (a) and (e) indicate the Brillouin zones of Ag(110) and SiNRs, respectively. The black arrows in (a)-(d) indicate band structures that originate from SiNRs. The bulk states and Shockley surface states of Ag(110) are indicated by ``BS'' and ``SS'', respectively.}
\end{figure*}

We then performed ARPES measurements to study the electronic structures of the SiNRs. The constant energy contours of SiNRs/Ag(110) and pristine Ag(110) are displayed in Figs. 2(a)-2(d) and Figs. 2(e)-2(h), respectively. Because the coverage of Si is less than one monolayer, there are remanent signals of the surface states of Ag(110) at the $\rm\bar{Y}$ points; this also indicates the cleanliness of our samples. On the Fermi surface, most bands derive from the Ag(110) substrate, except for two straight lines that are symmetric about the $\rm\bar{M}$--$\rm\bar{X}$--$\rm\bar{M}$ direction of Ag(110), as indicated by the black arrows in Figs. 2(a)-2(d). With increasing binding energy, the two parallel lines move close to each other [Fig. 2(b)] and merge into one line at $\sim$0.8 eV [Fig. 2(c)]. At higher binding energies, the two lines separate again, as shown in Fig. 2(d). This behavior indicates that these parallel bands have dispersion only in one direction, {\it i.e.}, along the SiNRs, which agrees well with their 1D character.

Notably, the bulk bands of Ag(110) have strong spectral weights near the $\rm\bar{X}$ point, where they overlap with the bands of the SiNRs. To disentangle the band structures of the SiNRs, we present ARPES spectra along Cuts 1-5 that are away from the $\rm\bar{X}$ point, as indicated in Fig. 3(a). Interestingly, we observe linearly dispersing bands with a Dirac cone-like shape, as shown in Figs. 3(b)-3(f). Based on our experimental results, the Fermi velocity is approximately 1.3$\times$10$^{6}$ m/s, which is slightly higher than that of graphene ($\sim$1.0$\times$10$^{6}$ m/s). The Dirac point is located 0.8 eV below the Fermi level, as indicated by the energy distribution curve in Fig. 3(g). Within our energy resolution, we did not observe an obvious band gap at the Dirac point, which suggests the existence of 1D massless Dirac fermions in the SiNRs. In addition, our ARPES spectra obtained using a different photon energy show negligible dispersion [Fig. 3(h)], which is consistent with the 1D nature of the bands. To better visualize the Dirac points, we present band structures along the $\rm\bar{M}$--$\rm\bar{X}$--$\rm\bar{M}$ direction, as shown in Figs. 3(i)-3(k). We observe a flat band at $E\rm_B\approx$0.8 eV that corresponds to the band degenerate points, which is absent in pristine Ag(111) [Fig. 3(l)]. We also performed scanning tunneling spectroscopy measurements to study the local density of states of the SiNRs, as shown in Supplementary Fig. S1. There is a dip at 0.8 eV below the Fermi level, which agrees with the Dirac point observed by ARPES. Notably, there are parallel bands with binding energy between 1.0 and 1.5 eV, as shown in Figs. 3(i)-3(k). These bands have strong photon energy dependence and might originate from the quantum well states of bulk band electrons.

\begin{figure*}[htb]
\centering
\includegraphics[width=16 cm]{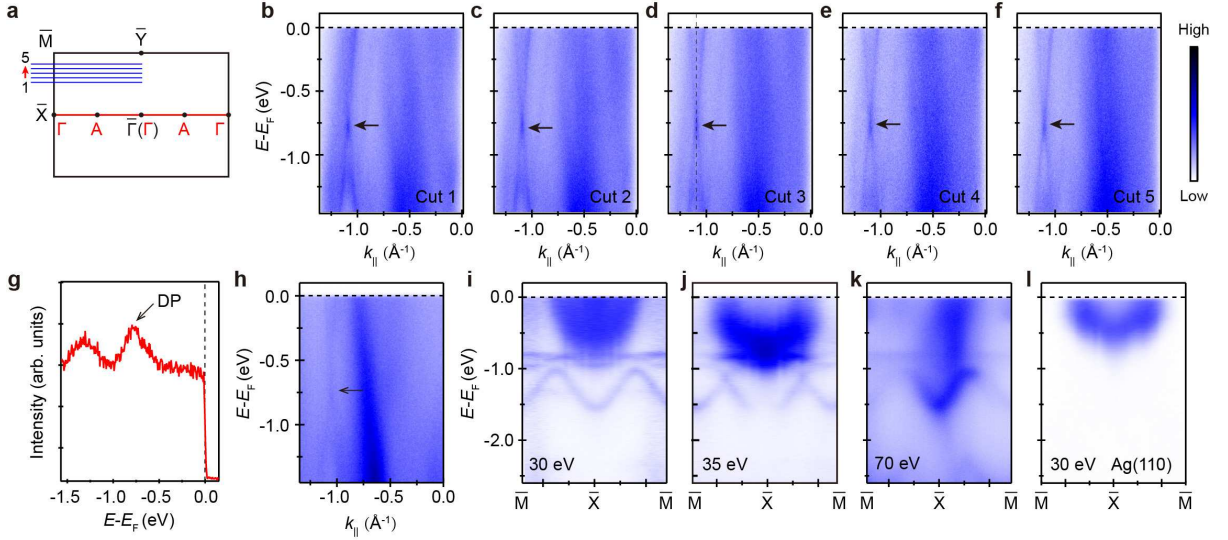}
\caption{{\bf ARPES measurements of the band structures.}(a) Schematic drawing of the Brillouin zones of Ag(110) (black) and SiNRs (blue). The blue lines indicate the five momentum cuts (Cuts 1-5) where (b)-(f) were taken. (b)-(f) ARPES intensity plots along Cuts 1-5 indicated in (a). (g) Energy distribution curve along the black dashed line in (d). The peak at $E\rm_B$=0.8 eV indicates the Dirac point. (h) ARPES intensity plot along Cut 1 obtained using 60 eV photons. (i) and (j) ARPES intensity plots along the $\rm\bar{M}$--$\rm\bar{X}$--$\rm\bar{M}$ direction of SiNRs/Ag(110) and pristine Ag(110), respectively. The flat band in (i) corresponds to the degenerate points.}
\end{figure*}

\section{III. First-Principles Calculations}

To understand our experimental results, we performed first-principles calculations including both SiNRs and the substrate. We take single-strand SiNRs as an example because double-strand SiNRs have similar electronic structures \cite{CerdaJI2016}. The structure model for calculation was constructed by placing a unit cell of SiNRs on a Ag(110)-(5$\times$2) surface slab with eight layers of Ag atoms. In the topmost Ag layer, the row of Ag atoms directly under the SiNRs were removed. Figures 1(c) and 1(d) show the relaxed structure model, which is in agreement with previous results \cite{CerdaJI2016}. Using the orbital-selective band unfolding technique \cite{MedeirosPVC2014,MedeirosPVC2015}, we unfolded the effective band structures of SiNRs/Ag(110) to the first BZ of Ag(110). For comparison with the ARPES spectra, the calculated band structures were projected onto the top two layers, {\it i.e.}, the nanoribbon and topmost Ag layer. The calculated band structure along the $\bar{\Gamma}$--$\rm\bar{X}$--$\bar{\Gamma}$ and $\rm\bar{M}$--$\rm\bar{X}$--$\rm\bar{M}$ directions of Ag(110) is shown in Figs. 4(a) and 4(b), respectively. Dirac cone-like bands exist at the $\rm\bar{X}$ point of Ag(110), with the Dirac point located at $\sim$0.75 eV, which is in qualitative agreement with our ARPES results. The other bands, which also have strong spectral weight, derive from the Ag(110) substrate (see Figs. S4 and S5 in the Supplementary Materials for more data). Along the $\rm\bar{M}$--$\rm\bar{X}$--$\rm\bar{M}$ direction, there is a flat band at $\sim$0.75 eV, which corresponds to the distribution of the Dirac points. Figures 4(c) and 4(d) show the calculated partial density of states (PDOS) of Si and Ag atoms, respectively. From the Fermi level to 1 eV below the Fermi level, where the Dirac cone is located, the density of states is dominated by the Si $p_z$ orbitals, as indicated by the green line in Fig. 4(c). Therefore, the Dirac cones mainly derive from the the SiNRs instead of the Ag(110) substrate.

\section{IV. Tight-Binding Model}

Next, we performed tight-binding (TB) analysis to understand the origin of the Dirac cones in the SiNRs. As discussed above, our first-principles calculations have confirmed that the Dirac cone mainly derives from the Si $p_z$ orbitals. Therefore, we only consider the $p_z$ orbitals of Si in the TB model. The Ag(110) substrate interacts with the SiNRs and modulates the hopping integrals between adjacent Si atoms. Based on the structure model of SiNRs/Ag(110), the Ag(110) surface has a missing row reconstruction along the [1\={1}0] direction, and the SiNRs sit directly above the missing rows. Therefore, there are two types of Si atoms with different heights and chemical environments: the lower Si atoms residing in the missing row troughs (Si$_{s}$) and the higher atoms leaning towards the short bridge sites at the top Ag rows (Si$_{ad}$) \cite{CerdaJI2016}. The TB Hamiltonian takes the form:

\begin{figure*}[htb]
\centering
\includegraphics[width=16 cm]{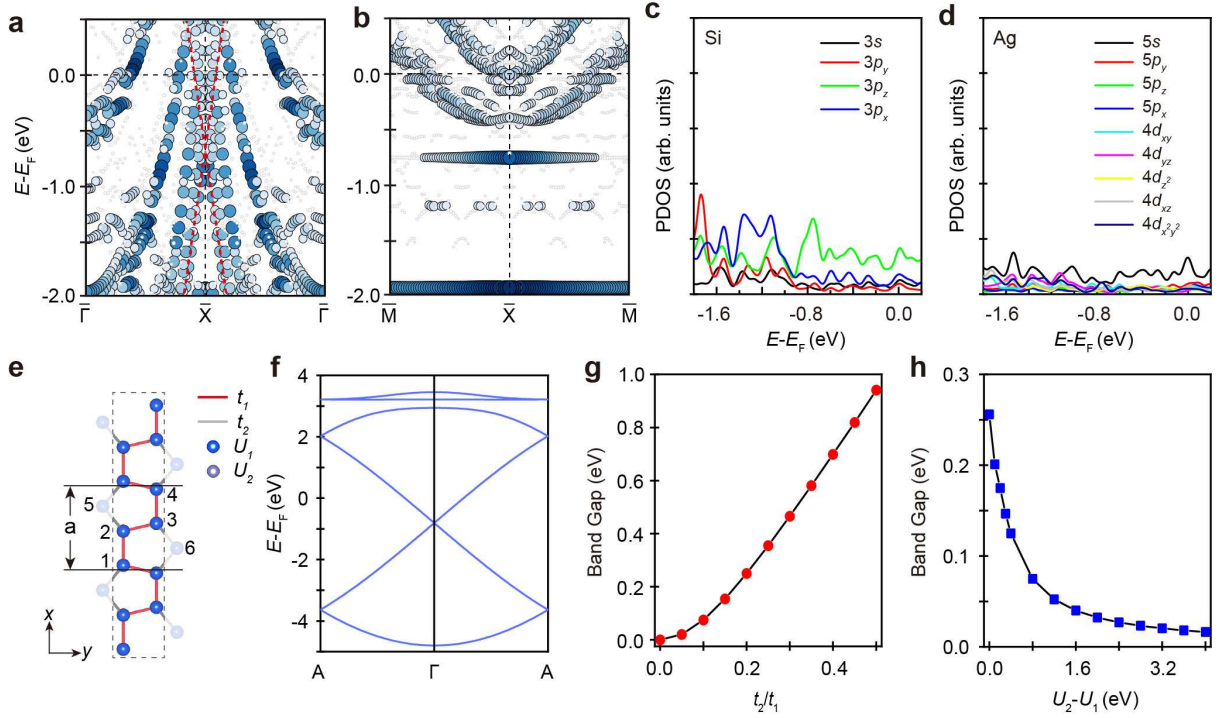}
\caption{{\bf First-principles calculations and tight-binding model analysis.} (a) and (b) Calculated band structures projected to nanoribbon and topmost Ag layer along $\bar{\Gamma}$--$\rm\bar{X}$--$\bar{\Gamma}$ and $\rm\bar{M}$--$\rm\bar{X}$--$\rm\bar{M}$, respectively. The red dashed lines indicate the Dirac cone at the $\rm\bar{X}$ point. (c) and (d) Calculated partial density of states of Si and Ag, respectively. (e) Schematic illustration of the TB model. There are six Si atoms in each unit cell, as indicated by the numbers 1 to 6. (f) Band structure calculated using the TB model. Parameters: $t_1$=2 eV, $t_2$=0.2 eV, $U_1$=-0.8 eV, and $U_2$=3.2 eV. (g) Band gap dependence on $t_2$/$t_1$. Parameters: $t_1$=2 eV, $U_1$=-0.8 eV, and $U_2$=0 eV. (h) Band gap dependence on $U_2$-$U_1$. Parameters: $t_1$=2 eV, $t_2$=0.2 eV and $U_1$=-0.8 eV.}
\end{figure*}

\begin{equation*}
\begin{split}
H=-t_1\sum\limits_{ij}^{n.n.}(a_i^\dag a_j+H.c.)+U_1\sum\limits_i a_i^\dag a_i\\-t_2\sum\limits_{ij}^{n.n.}(a_i^\dag b_j+H.c.)+U_2\sum\limits_i b_i^\dag b_i
\end{split}
\end{equation*}

where $a_i^\dag$ and $a_i$ are the creation and annihilation operators of the Si$_s$ 3$p_z$ orbitals, $b_i^\dag$ and $b_i$ are the creation and annihilation operators of the Si$_{ad}$ 3$p_z$ orbitals, $t_1$ is the hopping integral between the Si$_s$ 3$p_z$ orbitals, $t_2$ is the hopping integral between Si$_s$ 3$p_z$ and Si$_{ad}$ 3$p_z$, and U$_1$ and U$_2$ are the on-site energies imposed by the Ag(110) substrate on the Si$_s$ 3$p_z$ and Si$_{ad}$ 3$p_z$ orbitals. By fitting to the experimental Dirac point and Fermi velocity, we can obtain $t_1$=2 eV and $U_1$=-0.8 eV. These two parameters give rise to a 1D Dirac cone at the $\Gamma$ point, as shown in Fig. 4(f). Since the $\Gamma$ point of the SiNRs is equivalent to the $\rm\bar{X}$ point of Ag(110), our TB band structures agree well with our ARPES and first-principles calculation results. On the other hand, we realize that the Dirac cone is gapped when $t_2$ is nonzero. However, the gap quickly closes with decreasing $t_2$ or increasing $U_2$, as shown in Figs. 4(g) and 4(h). In the extreme case, the Dirac cone becomes gapless. Therefore, the gapless Dirac cones observed by our ARPES measurements indicate a small $t_2$ or a large $U_2$, which proves that Si$_s$ and Si$_{ad}$ are almost decoupled. This is reasonable since these two types of Si atoms are not coplanar and have different chemical environments. The Dirac cones mainly derive from Si$_s$ atoms with an armchair-like shape, as illustrated in Fig. 4(e). The Si$_{ad}$ atoms mainly contribute to the relatively flat bands above the Dirac cones.

As discussed above, the Dirac cones mainly derive from Si$_s$ atoms with an armchair-like shape. Therefore, we can simplify the TB model considering only Si$_s$ atoms to further understand the origin of the Dirac cone. The TB Hamiltonian takes the form:

\begin{equation*}
H=-t_1\sum\limits_{ij}^{n.n.}(a_i^\dag a_j+H.c.)
\end{equation*}

For simplicity, the on-site energy is set to be zero, which have no influence on our analysis results. The TB Hamiltonian can be transformed to the reciprocal space:

\begin{equation*}
H=-
  \begin{pmatrix}
     0 & t_1e^{ikl_1} & 0 & t_1e^{-ikl_2}\\
     t_1e^{-ikl_1} & 0 & t_1e^{ikl_2} & 0\\
     0 & t_1e^{-ikl_2} & 0 & t_1e^{ikl_1}\\
     t_1e^{ikl_2} & 0 & t_1e^{-ikl_1} & 0\\
  \end{pmatrix}
\end{equation*}

where $l_1$ and $l_2$ are the two kinds of bond lengths projected along the periodic direction. At the $\Gamma$ point, we can obtain the eigenvalues of the Hamiltonian: $E_1$=-2$t_1$, E$_{2,3}$=0 and $E_4$=2$t_1$. Obviously, there is a twofold band degeneracy corresponding to the Dirac point. We then take the spin degree of freedom into account. Because of the presence of inversion symmetry and time-reversal symmetry, each band of the SiNRs are spin degenerate. Therefore, there exists a fourfold Dirac point at the $\Gamma$ point. Notably, the armchair-like Si chain is equivalent to the SSH model because of the alternating short and long bonds along the periodic direction, which is the reason for the emergence of topological band structures. By adjusting the hopping integrals between atoms, the Dirac cone can be gapped and become an 1D topological insulator or trivial insulator.

\section{V. Conclusion}

In summary, we report the observation of 1D Dirac fermions in SiNRs based on ARPES measurements, first-principles calculations, and tight-binding model analysis. The Dirac bands emerge because of the unique pentagonal nature of SiNRs and can be described by the simple SSH model. Our results provide new opportunities to study and manipulate the rich physical properties of 1D Dirac materials. For example, by applying external magnetic field, the time-reversal symmetry will be broken, which will result in a topological phase transition in SiNRs: the 1D Dirac cone will split into two pairs of 1D Weyl cones (see Supplementary Materials for a brief theoretical analysis). These intriguing properties remains to be explored both theoretically and experimentally. Finally, we would like to mention that SiNRs interact strongly with the Ag(110) substrate. Although the conducting substrate has little influence on the fabrication of optical and plasmonic devices, it impedes transport measurements on the exotic properties of 1D Dirac fermions. Therefore, we hope that our work could stimulate further research efforts to prepare SiNRs on insulating substrates or search for freestanding 1D Dirac materials.

\section{VI. Methods}

Single-crystal Ag(110) was cleaned by Ar ion sputtering and annealing cycles. The cleanliness of the Ag(110) surface was confirmed by LEED and ARPES measurements. SiNRs were prepared by evaporating silicon from a Si wafer onto Ag(110). The flux of silicon was kept at 0.08-0.1 ML/min. ARPES experiments were performed at Beamlines BL-1 \cite{IwasawaH2017} and BL-9A of the Hiroshima synchrotron radiation center. The samples for ARPES measurements are predominantly double-strand SiNRs, and the area ratio of single- and double-strand SiNRs is approximately 1:25. During ARPES measurements, the pressure was $\sim$1.0$\times$10$^{-9}$ Pa and the temperature of the samples was kept at $\sim$30 K. STM experiments were carried out in a home-built low-temperature STM and all STM images were taken at $\sim$77 K.

First-principles calculations based on density functional theory (DFT) were performed with Vienna ab initio simulation package (VASP) \cite{KresseG1993,KresseG1996}. The projector-augmented wave pseudopotential \cite{BlochlPE1994} and Perdew-Burke-Ernzerhof exchange-correlation functional \cite{PerdewJP1997} were used. The energy cutoff of plane-wave basis was set as 250 eV, and the vacuum space was set to be larger than 15 \AA. The first BZ was sampled according to the Monkhorst-Pack scheme. We used a $k$ mesh of 1$\times$3$\times$1 for structural optimization and 3$\times$7$\times$1 for the self-consistent calculations. The positions of the atoms were optimized until the convergence of the force on each atom was less than 0.01 eV/\AA. The convergence condition of electronic self-consistent loop was 10$^{-6}$ eV. The Ag atoms near the Si atoms were counted in the calculation of the partial density of states. The band structure of the SiNRs/Ag(110) system was calculated using the orbital-selective band unfolding technique \cite{MedeirosPVC2014,MedeirosPVC2015}.

\section{VII. Associated Content}
{\bf Supporting Information}\\
Topological Phase Transition in the Presence of Magnetic Field; Scanning Tunneling Spectroscopy (STS) Measurements; Band structures of Ag(110); Orbital Resolved Band structures of SiNRs on Ag(110); First-Principles Calculation Results of Double-Strand SiNRs on Ag(110)

\section{VIII. Acknowledgement}
We acknowledge Prof. J. Ma for fruitful discussions. This work was supported by the Ministry of Science and Technology of China (Grant No. 2018YFE0202700), the National Natural Science Foundation of China (Grants No. 11974391, No. 11825405, No. 1192780039, and No. U2032204), the Beijing Natural Science Foundation (Grant No. Z180007), the International Partnership Program of Chinese Academy of Sciences (Grant No. 112111KYSB20200012), and the Strategic Priority Research Program of Chinese Academy of Sciences (Grants No. XDB33030100 and No. XDB30000000). ARPES measurements were performed under the Proposal No. 19AG006 and 19BG028. We thank the N-BARD, Hiroshima University for supplying liquid He.

S.Y., H.Z., Y.F., and Y.W. contributed equally to this work.

\section{IX. References}

\end{document}